# Subspace orthogonalization as a
# mechanism for binding values to space

W. Jeffrey Johnston[1*+], Justin M. Fine[2*+],
Seng Bum Michael Yoo[3,4], R. Becket Ebitz[5], and
Benjamin Y. Hayden[2]

1. Center for Theoretical Neuroscience and Mortimer B. Zuckerman Mind, Brain, and Behavior Institute, Columbia University, New York, New York
2. Department of Neuroscience, Center for Magnetic Resonance Research, and Department of Biomedical Engineering, University of Minnesota, Minneapolis, MN 55455
3. Department of Biomedical Engineering, Sunkyunkwan University, and Center for Neuroscience Imaging Research, Institute of Basic Sciences, Suwon, South Korea, Republic of Korea, 16419
4. Current address: Department of Brain and Cognitive Sciences, Massachusetts Institution of Technology, Cambridge, Massachusetts, MA, 02139
5. Department of Neuroscience, Université de Montréal, Montréal, Quebec, Canada

* Equal contribution
+ Correspondence: wjeffreyjohnston@gmail.com, justfineneuro@gmail.com

**Funding statement**
    This research was supported by a National Institute on Drug Abuse Grant R01 DA038615 (BYH), MH124687 (BYH), NSF 1707398 (WJJ), Simons Foundation 542983SPI (WJJ), Gatsby Charitable Foundation GAT3708 (WJJ)

**Competing interests**
    The authors have no competing interests to declare.

**Acknowledgements**
    We thank Bill Vinje, who led an excellent summer journal club on the binding problem at UC Berkeley in 2001. We thank Maya Wang, Tyler Cash-Padgett, Marc Mancarella, Caleb Strait, Tommy Blanchard, and Brianna Sleezer for assistance with data collection. We also thank Allison Ong and Aleyna Silcott for administrative support.

**ABSTRACT**

When choosing between options, we must solve an important binding problem. The values of the options must be associated with information about the action needed to select them. We hypothesize that the brain solves this binding problem through use of distinct population subspaces. To test this hypothesis, we examined the responses of single neurons in five reward-sensitive regions in rhesus macaques performing a risky choice task. In all areas, neurons encoded the value of the offers presented on both the left and the right side of the display in semi-orthogonal subspaces, which served to bind the values of the two offers to their positions in space. Supporting the idea that this orthogonalization is functionally meaningful, we observed a session-to-session covariation between choice behavior and the orthogonalization of the two value subspaces: trials with less orthogonalized subspaces were associated with greater likelihood of choosing the less valued option. Further inspection revealed that these semi-orthogonal subspaces arose from a combination of linear and nonlinear mixed selectivity in the neural population. We show this combination of selectivity balances reliable binding with an ability to generalize value across different spatial locations. These results support the hypothesis that semi-orthogonal subspaces support reliable binding, which is essential to flexible behavior in the face of multiple options.

# INTRODUCTION

Humans and non-human animals are adept at choosing between offers based on their values. When choosing, decision-makers must link the estimated values of each offer to features that can be used to select it, such as the actions needed to choose the preferred option (Kable and Glimcher, 2009; Samejima et al., 2005; Rangel et al., 2008; Wunderlich et al., 2009). To choose between two items, then, we must keep in mind at least four pieces of information - two values and two actions, and then appropriately bind each value to its corresponding action (Hayden and Moreno-Bote, 2018). Failure to bind these elements correctly will lead to an erroneous assignment of value to action and a suboptimal choice. The problem of how the brain binds value to action remains unresolved, but is fundamental to decision-making in the neuroscience of economic choice and in general (Cai and Padoa-Schioppa, 2014; Hare et al., 2011; Hayden and Moreno-Bote, 2018;  Knudsen and Wallis, 2022; Padoa-Schioppa and Assad, 2008;  Rangel et al., 2008; Rosech et al., 2009; Shadlen and Movshon, 1999;Wunderlich et al., 2009; Yin et al., 2019; Stoet and Hommel, 1999).

Much recent research highlights the information coding power and flexibility of neural populations (Chung and Abbott, 2021; Ebitz and Hayden, 2021; Elsayed et al., 2016; Mante et al., 2013; Urai et al., 2021; Saxena and Cunningham, 2019; Vyas et al., 2020). Significant flexibility of population codes is attributable to the fact that individual neurons often respond to linear and nonlinear mixtures of multiple task-relevant features (i.e., mixed selectivity; Blanchard et al., 2018; Fusi et al., 2016; Hocker et al., 2021; Raposo et al., 2014; Rigotti et al., 2013). This mixed selectivity naturally emerges from the random projection of an input representation (which may be low-dimensional) to a population of nonlinear neural units (Barak et al., 2013; Babadi and Sompolinsky, 2014; Rigotti et al., 2013). The resulting representation is often higher dimensional than the input and provides a representation that downstream regions can use to discriminate between different combinations of task-relevant features that may not have been discriminable in the input (Babadi and Sompolinsky, 2014; Barak et al., 2013; Fusi et al., 2016; Ganguli and Sompolinsky, 2012). We hypothesize that neural systems solve the binding problem by encoding the value of different task variables along semi-orthogonal dimensions of the full population space. We refer to these lower-dimensional spaces embedded in the full population space as subspaces. In particular, we propose that core reward regions represent different feature combinations (i.e., different offer values presented at different spatial positions, in our case) in distinct subspaces.

How could the brain leverage mixed selectivity to form distinct subspaces within the same underlying neural population? Population codes that rely on nonlinear mixed selectivity naturally produce distinct subspaces (Fusi et al., 2016; Parthasarathy et al., 2017). The creation of subspaces through nonlinear mixing enhances the separability of different patterns of bound features, increasing the reliability reading out these feature combinations with minimal confusability (Johnston et al., 2020; Litwin-Kumar et al., 2017; Rigotti et al., 2013). Such high-dimensional coding schemes are often disadvantaged by brittleness to noisy inputs and a reduced ability to generalize across different contexts (Barak et al., 2013; Fusi et al., 2016; Renart and Machens, 2014). In contrast, linear mixed selectivity provides a low-dimensional and factorized representation that supports generalization across different contexts (Bernardi et al., 2020). These different advantages of the two forms of selectivity underlie competing theories about whether neural population codes are expected to have high- or low-dimensionality (Bernardi et al., 2020; Flesch et al., 2022; Cueva et al., 2020; Gallego et al., 2017; Jazayeri and Ostojic, 2021; Rigotti et al., 2013). Previous studies of neural codes, removed from questions related to binding, observed that subspaces were seldom fully orthogonal or factorized; instead, they tend to blend elements that increase dimensionality and orthogonality (non-linear mixed selectivity) or decrease it (linear mixed selectivity, Bernardi et al., 2020; Blanchard et al., 2018; Fusi et al., 2016; Parthasarathy et al., 2017; Raposo et al., 2014; Urai et al., 2022). Blended codes reside in the middle of the continuum between subspace orthogonality and collinearity. Therefore, we asked how the mixing of linear and nonlinear codes could support a neural geometry with subspaces that ensure separability and binding alongside generalizability during value-based choice.

Here, we examined the geometry of neural population representations in five brain areas associated with reward: the orbitofrontal cortex (OFC 11/13), the pregenual (pgACC 32) and posterior cingulate (PCC 29/31) cortices, the ventromedial prefrontal cortex (vmPFC 14) and the ventral striatum (VS). All regions were recorded in an identical risky choice task involving two risky options separated in space (left and right of a central fixation point, Strait et al., 2014). We found that neurons in all five regions show clear signatures of both linear and nonlinear mixed selectivity and use semi-orthogonal and therefore separable subspaces for both choice options appearing in opposing spatial locations. To understand the separate contributions of both kinds of selectivity, we developed a mathematical framework that links subspace orthogonality, binding, and generalization. Using this framework, we demonstrate that these representations – which blend linear and nonlinear mixed selectivities – produce a code that results in both a low misbinding error rate (i.e., the subspaces are sufficiently orthogonal and high-dimensional) and – despite this – that preserves compositionality of value, which is important for learning and generalization. Critically, we found that sessions with more orthogonal subspaces for the two offers also tended to have fewer suboptimal choices, which is consistent with our theory: more orthogonal subspaces will more reliably bind the offers to their locations and reduce the rate of any misbinding errors. We also find that, despite this, the representations for value remain reliably abstract and generalizable. These results provide support to the hypothesis that the brain makes use of subspace orthogonalization to solve the important binding problem associated with choice. While they relate directly to important questions in neuroeconomics, we anticipate that these ideas may help explain how the brain implements binding more generally.

## RESULTS

**Subjects show behavior consistent with task understanding in risky choice task**

We analyzed data collected from six rhesus macaques trained to perform a two-option **asynchronous gambling task** that we have used several times in the past (e.g., Strait et al., 2014, **Figure 1**). Behavioral data were consistent with patterns we have previously reported (for the most complete analyses, see especially Farashahi et al., 2018 and 2019). Briefly, on each trial, subjects use a saccade to select one of two risky offers. The two offers are presented in sequence and on different sides of a computer screen (left or right). Each option appears for 400 ms, followed by a blank screen for 600 ms (thus, there is a one-second stimulus onset asynchrony between the two offers). Then, the two offers reappear simultaneously and the animal makes a choice by fixating their preferred offer for 200 ms. Each offer is defined by a probability (0-100%, 1% increments) and stakes (large or medium reward, 0.240 and 0.165 mL juice, respectively). Thus, once the animal selects an offer, the probability of the offer dictates how likely they are to receive any reward and the stakes determine the size of that possible reward. The probabilities and stakes associated with both offers are chosen randomly for each option. The order of presentation (left first vs. right first) is randomized by trial.

All six animals were more likely to choose offers with a higher expected value – i.e., the product between probability and stakes. As in past studies using this task, all subjects were reliably risk-seeking (Heilbronner and Hayden, 2013). For our analyses here, we estimated the subjective value underlying each animal's choices. In particular, while the nominal expected value is the product of the probability and stakes (as mentioned above), the different subjects may differ in how they weigh these two variables. Indeed, the best fitting model we considered incorporates non-linear weightings of probability and stakes, and predicts the animal's choice with 87 - 95% accuracy. The subjective value model is fit independently for each session. In the rest of the manuscript, we use the term *value* to mean the model-derived subjective value for the session from which it was collected.

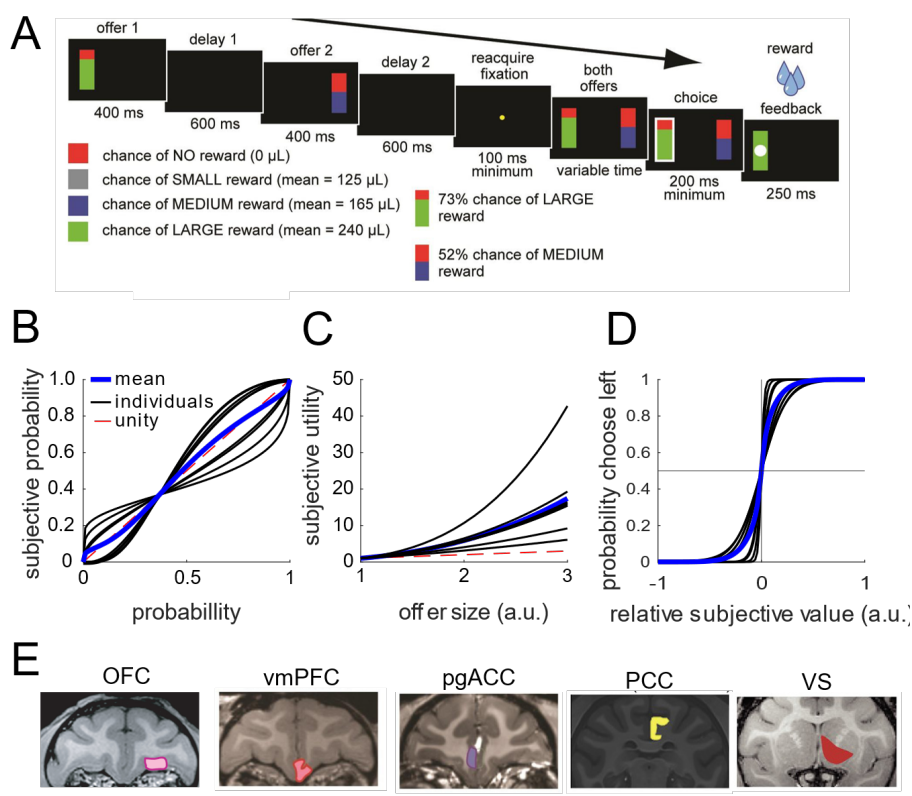

**Figure 1.** Task outline and brain areas. **A.** The risky-choice task is a sequential offer decision-task that we have used in several previous studies (e.g., Strait et al., 2014). In the first 400 ms, subjects see the first offer as a bar presented on either the left or right

side. The above shows an example on the left. This offer is followed by a 600 ms delay, a 400 ms offer 2 window, another delay (2) window, and then choice. The full task involved either small, medium or large reward offers. The small reward trials were actually those with safe (guaranteed) offers.  We only analyzed the risky choice trials, which were those including the medium and large reward. **B-D.** From left to right, the plots show for all subjects the model fitted (left) subjective probability, (middle) subjective utility, and (right) relative subjective value choice curves.  **E.** MRI coronal slices showing the 6 different core reward regions that were analyzed.

**Hypothesis: value is bound to space by nonlinear codes**

We hypothesize that the core value regions we studied bind offer value to spatial position using a nonlinear population code. In particular, we predict that the value of the left and right offers will be encoded in distinct (not collinear but not necessarily orthogonal) subspaces, which we refer to as *subspace binding*. Suppose we have a perfectly factorized representation of offer value and position. In other words, the two variables are encoded in the firing rates of a population of neurons, and the response of a neuron to a particular combination of value and space is a linear function of its response to each variable individually (e.g., a weighted sum). For a linear representation of value, this would mean that the representation of value and space together would manifest as a rectangular structure in neural population space (schematized in **Figure 2D**, top). This coding scheme does not bind value to position. Formally, a maximum likelihood decoder would find two equally likely interpretations of the population response to the two offers: it would be able to decode the two offer values and the two offer positions, but not *which* offer value was at which position. Importantly, this ambiguity exists even for a nonlinear representation of value.

Suppose instead that there is a nonlinear interaction between the encodings of the two variables. This encoding could be conjunctive – such that a particular neuron responds strongly to a particular combination of value and position and weakly to all other combinations. Importantly, a conjunctive response cannot be explained as a weighted sum of responses to value and position individually (in a linear model, this manifests as an interaction term between variables). Such an encoding scheme binds value to position and breaks the ambiguity for decoding described above. Such a nonlinear interaction will have the effect of pushing the representations of the left and right offers into distinct population subspaces (see **Methods** for more detail on and mathematical formalization of this hypothesis). These subspaces serve as distinct parts of high dimensional space where different features are bound together.

**Single neurons have nonlinear responses to different combinations of value and space**

We hypothesized that we would observe evidence for this orthogonalizing process in brain regions that represent the offers in our task. To test this prediction, we analyzed the responses of 929 neurons across five brain regions, with two subjects per region (**Figure 1**; n=242 neurons in OFC, n=156 in vmPFC, n=255 in pgACC, n=152 in PCC, and n=124 in VS). We observed diverse responses from neurons in each brain region (**Figure 2A**). These diverse responses also give rise to diverse value-response functions (**Figure 2B**), many of which are consistent with our hypothesis of a nonlinear interaction between value and space (e.g., the example neurons from OFC and pgACC).

To investigate whether these neural responses were best explained by linear or nonlinear interactions between representations of value and space, we fit several linear regression models. First, we fit a model that explains the neural responses only in terms of noise (**Figure 2C**, "noise"), then a model with terms for value and space and only a linear interaction (**Figure 2C**,

"linear"), and, finally, a model with terms for value and space as well as a nonlinear interaction between them (**Figure 2C**, "interaction"). For both the linear and interaction models, we fit versions of the model with both linear and spline-based representations for value (see **Methods** and **Figure S1** for more details). Then, we compared the fits of these models to the data, through a Bayesian model stacking analysis based on approximate leave-one-out cross validation (see **Methods** and Yao et al., 2017; **Figure 2C**). In every brain region, we found a substantial proportion of neurons whose responses were best fit by the nonlinear interaction model (OFC: 12%, PCC: 19%, pgACC: 10%, vmPFC: 12%, VS: 13%). These results are not predicted by conventional "neuron doctrine" approaches, but are predicted by our subspace binding hypothesis, since only the interaction models can produce non-parallel subspaces for the value of left and right offers. Thus, this analysis already provides support for our main hypothesis.

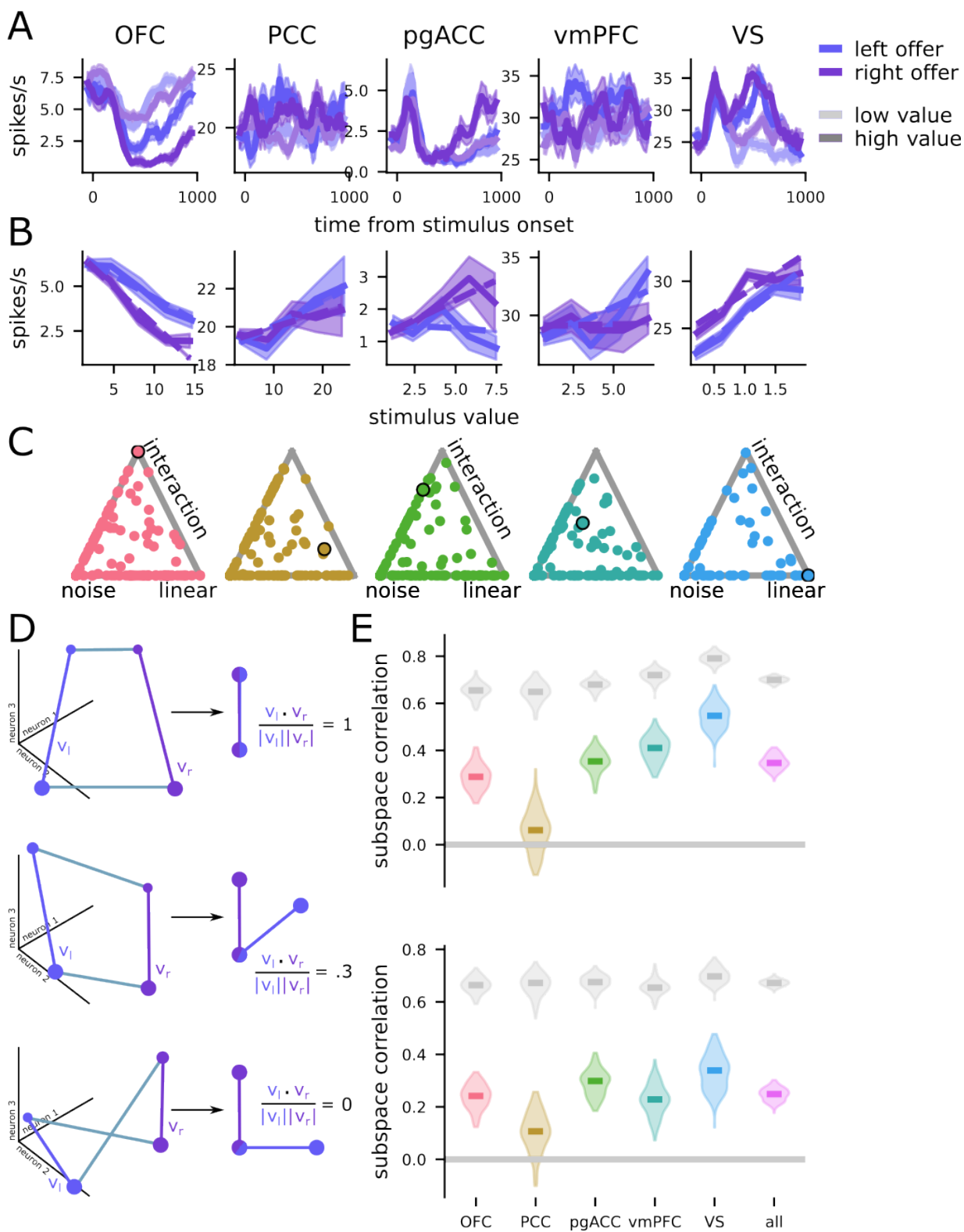

**Figure 2.** Example neurons, model comparison, and subspace correlations. **A.** The firing rates of example neurons from each region during the offer window, shown for high and low value offers presented on the left or right side (100 ms boxcar filter,

shaded area is SEM). **B.** The value-response function for each neuron in **A**. The value-response function fit by the linear regression model with an interaction term is overlaid (dashed lines). **C.** A simplex showing the weight given to each of the noise-only, linear, and interaction regression models by the Bayesian model stacking analysis. The points corresponding to the example neurons shown in **A** and **B** have dark outlines here. Both the linear and interaction categories include both linear and spline value representation models. **D.** Schematic of three different representational geometries that would lead to different subspace correlation results. (top) Two perfectly aligned value vectors $v_l$ and $v_r$ in population space (left) would produce a subspace correlation close to 1 (right). (middle) Two partially aligned value vectors $v_l$ and $v_r$ in would produce a subspace correlation between 0 and 1 (note there is an additional possibility: partially aligned but negatively correlated subspaces; not schematized). (bottom) Two unaligned value vectors $v_l$ and $v_r$ would produce a subspace correlation close to 0. **E.** Subspace correlations for all regions for the offer presentation window. The gray point is the subspace correlation expected if the left- and right value subspaces were aligned and corrupted only due to noise. **E.** Same as **D.** for the delay period.

**Neural populations use separable but partially overlapping subspaces for different offers**

To assess the degree of overlap between the left- and right-value subspaces at the population level, we use the coefficients from the regression model described in the previous section. These models describe a representation of the offer value for each presentation side during the offer presentation and delay task epochs. We want to know how similar this representation of value is across the different offer positions. To do this, we define the value-encoding subspace for each offer side, which is the vector between low- and high-value offers presented on the left ($v_l$) and right ($v_r$) sides (**Figure 2D**; and see **Figure S1** for a replication of these results when the subspaces are not constrained to be vectors). We then quantify the similarity between the two value representations by taking the correlation (i.e., unit vector dot product) between the two vectors.

What would different levels of subspace correlation mean for our subspace binding hypothesis? A subspace correlation close to 1 (or, more specifically, the noise ceiling, shown in gray in **Figure 2E-F**) indicates primarily linear interactions between the offer value and position encodings (**Figure 2D**, top; the two value vectors $v_l$ and $v_r$ are parallel to each other). Such a result would indicate that the coding scheme does not implement subspace binding. Conversely, subspace correlations substantially less than 1 (and also less than the noise ceiling in **Figure 2E-F**) indicate nonlinear interactions between offer value and position codes, and are consistent with our hypothesis of subspace binding. The subspace correlation could take on an intermediate value (i.e., the two value vectors $v_l$ and $v_r$ point in the same direction, but are not fully parallel and not fully orthogonal; **Figure 2D**, middle). This would mean that there is a mixture of both linearly and nonlinearly interacting value and position representations. The intermediate level of subspace correlation (i.e., between 1 and 0) both supports binding and has a potentially important additional benefit: it allows for the generalization of the value code across offer positions and epochs. We discuss this benefit more below. Other possibilities that our analysis approach could detect would include no subspace correlation (**Figure 2D**, bottom; the two value vectors $v_l$ and $v_r$ are fully orthogonal), or negative subspace correlation (not schematized).

We first computed the subspace correlation between left and right offers during the 400 ms offer presentation window. We find that in all five brain regions, the subspaces for left and right offers are significantly less correlated than would be expected if the two underlying value representations were the same and the imperfect correlation arose only from noise in neural firing rates ($p < 0.001$ in all cases, **Figure 2E**; and see **Methods** for reliability control, cf. Kimmel et al., 2020). In other words, we found that subspaces were more orthogonal than we would expect if the code for value were the same across spatial positions (i.e., if value were perfectly factorized from spatial position). In other words, these neurons exhibit the surprising, and to our knowledge never before reported property that their response to the value of an offer changes in a nonlinear way depending on where in space that offer appears. This difference can be used to solve the value-space binding problem, and we provide a theoretical prediction for the reliability of that solution (**Figure 3**).

Next, we asked whether the above subspace separation also carried into the delay period. Indeed, we found similar results when repeating the above comparison for the delay period (all $p$ < 0.001, **Figure 2F**). These results indicate that the population codes for distinct value-space and value-time combinations reside in four partially distinct population subspaces.

So far, we have shown evidence that value is not encoded in a single common subspace. One possible reason for this would be that there is an independent code for value at each time point and in each spatial position. In that case, we would expect the subspaces described above to be mutually orthogonal – that is, to have correlations with each other that are close to zero. However, we found subspace correlations were greater than zero ($p$ < 0.001) across all regions except PCC in both the offer presentation and delay windows (**Figure 2E and F**). In sum, the fact that correlations are greater than 0 but less than 1 indicates that the representational geometry is at a middle point between a single common value subspace, and totally distinct value subspaces for each offer position. Thus, the data are consistent with our hypothesis that binding is achieved by the representation of different offer values in different subspaces. The data also point to something further: that the code may not fully give up the putative benefits of a common value code, which include generalization (Bernardi et al., 2020; Johnston & Fusi, 2022) and rapid learning (Dosher & Lu, 2017; van Steenkiste et al., 2019). We quantify how this code supports generalization below (**Figure 3F**).

**Degree of subspace orthogonality predicts the rate of suboptimal choices**

If separable value-space subspaces serve to implement binding, then we would expect that less separated subspaces would be associated with less reliable binding. While we cannot directly identify binding errors in this experimental design, we can see their effects on behavior: binding errors necessarily reduce the rate at which the animals select the offer with higher value. In particular, we would expect that if reduced subspace correlation reflects more faithful binding, then increased subspace correlation should be positively associated with suboptimal choices (i.e., the rate at which the lower value offer is selected). Because this analysis requires simultaneous recordings from a large number of single neurons, we focused on sessions with at least thirty simultaneous neurons; this criterion resulted in two subjects in OFC, one in pgACC, and two in PCC.

We analyzed the subspace correlations for trial epochs that the animal actually compares to inform a choice. Namely, this is subspace correlations for the 1st and 2nd offers within a trial, and is broken up into comparing the value subspace for offer 1 presented on the left side and offer 2 on the right side as well as offer 1 presented on the right side and offer 2 on the left side. As above, we performed the analysis separately for the offer-on and delay periods. The proportion of suboptimal choices across all subject sessions examined ranged from 13% to 18%. We found that the correlation between the proportion of suboptimal choices and subspace correlations was significantly positive during the delay period for both offer sequences (i.e., left first or right first; permutation test, OFC: mean both $r = 0.21$ , $p = 0.009$; pgACC: $r = 0.15$, $p = 0.013$; PCC: mean both $r = 0.18$, $p = 0.011$). Interestingly, this same quantity was not significant during the offer-on period (all $p > 0.05$). Together, these results indicate that increases in separability between the two offer value subspaces were indeed linked to improved choice accuracy.

**Semi-orthogonal subspaces offer both binding and generalization**

Above, we showed that the subspace correlation between value representations for offers presented at different locations and at different times are between the two extremes of perfect correlation and full orthogonality. That is, the value subspaces are semi-orthogonal. Semi-orthogonal representations of the same stimulus feature in different contexts have been observed in many other studies as well (e.g., Elsayed et al., 2016; Flesch et al., 2022; Tang et al., 2020; Yoo and Hayden, 2020). Here, we investigate this property of neural representations in more detail by developing an analytic theory that links this semi-orthogonality to the reliability of both the binding of value to space (or to time) and the ability to generalize a decoder learned for offers on one side (or offers presented at one time) to the other side (or time). To simplify this theory as well as allow for the use of standard linear decoder techniques (as used in **Figure 3F**), we binarize the continuous expected value variable into two categories: high and low. Thus, we consider representations that consist of four points; for space, these are high value offers presented on the left, high value offers presented on the right, and similarly for low value offers. We also consider an analogous set of four points divided across different presentation times, replacing left and right with offer 1 and offer 2.

Then, we decompose the resulting geometric structure (i.e., the particular arrangement of the four points in population space) into two components: First, a rectangular scaffold, where one axis of the rectangle corresponds to offer value and the other corresponds to offer position (**Figure 3A**, yellow lines), and, second, high-dimensional perturbations applied to that scaffold for each of the four points (**Figure 3A**, dark purple lines). We refer to the length of the rectangular value axis as the linear distance (**Figure 3A**, $d_L$) and the length of the high-dimensional perturbations as the nonlinear distance (**Figure 3A**, $d_N$). Then, we relate these linear and nonlinear distances to the subspace correlation (**Figure 3A**, bottom). In particular, a large linear distance and small nonlinear distance implies a high subspace correlation (**Figure 3A**, bottom left); similar linear and nonlinear distances imply a moderate subspace correlation (**Figure 3A**, bottom center); and a large nonlinear distance and small linear distance implies a low subspace correlation (**Figure 3A**, bottom right).

Next, our main theoretical result is to relate these linear and nonlinear distances to the expected binding and generalization error rates that would be made by a downstream decoder reading out from these neural populations. First, we investigate how the binding error rate depends on these distances along with the noise present in the neural responses. We show that binding errors only depend on the nonlinear distance and the noise level of the representation (**Methods**). This means that the binding error rate can be low even when subspace correlations are relatively high, so long as the nonlinear distance is larger than the magnitude of noise (**Figure 3B**).

Second, we investigate how the generalization error rate depends on the linear distance, nonlinear distance, and the noise. In particular, we derive a predicted error rate for a linear decoder trained to decode the value of left offers when it is applied to offers presented on the right (and vice versa). This measure of generalization is often referred to as the cross-condition generalization performance (Bernardi et al., 2021). We show that increases in linear distance will decrease the rate of generalization errors, while increases in nonlinear distance will increase this rate (**Methods**). Thus, low subspace correlations are unlikely to yield low generalization error rates (**Figure 3C**). Because the subspace correlations we measured were between zero and one (**Figure 2E-F**), we hypothesize that the neural representations will be consistent with both low binding error rates and low generalization error rates.

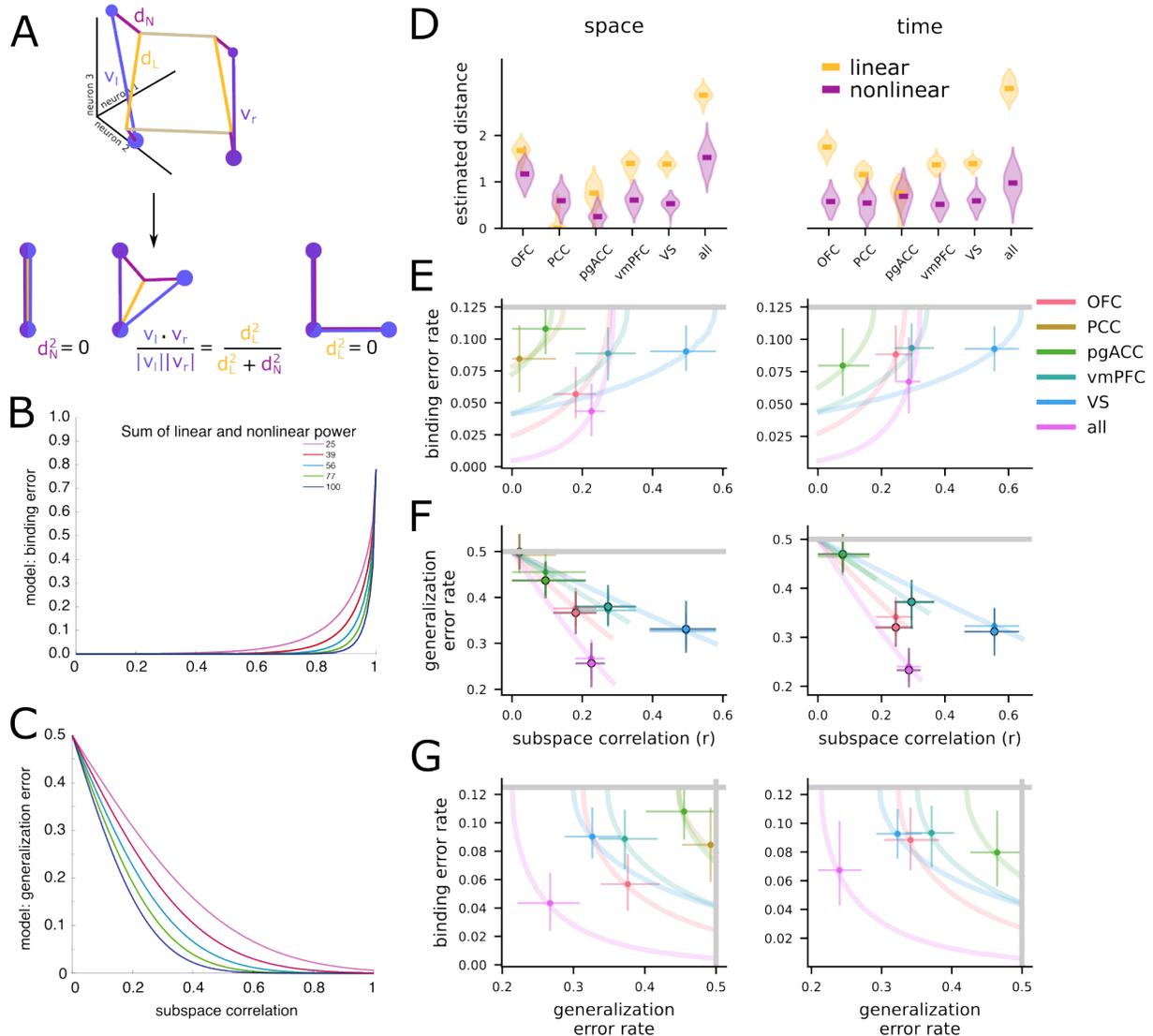

**Figure 3.** Formalizing subspace structure through a geometric theory of binding and generalization for neural codes. **A.** Schematic of the geometric decomposition. (top) The representation from **Figure 2** is decomposed into linear $d_L$ (yellow) and nonlinear $d_N$ (purple) components. (bottom) The relative length of these components determines the subspace correlation from before: $d_N = 0$ and $d_L > 0$ implies perfect subspace correlation (bottom left), both $d_N > 0$ and $d_L > 0$ implies intermediate subspace correlation (bottom middle), and $d_N > 0$ while $d_L = 0$ implies zero subspace correlation (bottom right). **B.** The relationship of the binding error rate predicted by our theory with subspace correlation.

The different lines are codes with different sums of squared linear and nonlinear distances. The line is created by varying the tradeoff between linear and nonlinear distance such that the sum remains constant. The left side of the line is when linear distance is zero and nonlinear distance is the total distance; the right side is the opposite extreme. **C.** The same as **B** but for the generalization error rate. **D.** (left) The nonlinear and linear distances estimated for the left and right value codes within each brain region. (right) The same as on the left, but for the offer 1 and offer 2 value codes. The violin plot shows the distribution of bootstrap resamples. **E.** The predicted binding error rate as a function of subspace correlation for each region, derived from the distance estimates in **D**. The left-right plot convention is the same as in **D**. The gray line shows the chance level of binding errors. **F.** The predicted generalization error rate as a function of subspace correlation for each region, derived from the distance estimates in **D** (for the open circles) and computed empirically with a linear decoder (for the outlined circles). The gray line is the chance level. The left-right plot convention is the same as in **D**. **G.** Each region shown on the plane defined by the generalization and binding error plane, derived from the distance estimates in **D**. The gray lines are the chance levels for each of the error types. The left-right plot convention is the same as in **D**.

**Value representations are consistent with low binding and generalization error rates**

Now, we apply the theoretical framework developed in the previous section to our experimental data. First, we estimate the distances between all pairs of the four points using an unbiased and cross-validated distance metric (see **Methods**), which avoids the upward bias common to many other methods of distance estimation. Then, we decompose the resulting distance matrix into linear and nonlinear components, as defined above (see **Methods**). We apply this analysis separately to the left- and right-value code as well as the offer 1- and offer 2-value code. For the space-value code, we find significant nonlinear distances for all brain regions except pgACC and significant linear distances for OFC, vmPFC, and VS, but not PCC and pgACC (**Figure 3D**, left). For the time-value code, we find significant nonlinear distances for all regions and significant linear distances for all regions except pgACC (**Figure 3D**, right).

Then, we combine these nonlinear distance estimates with the theory developed in the previous section to produce a predicted rate of binding errors. In agreement with the distance estimates and because the rate of binding errors only depends on the nonlinear distance, we find binding error rates significantly below chance for all regions except for the space-value code in pgACC (**Figure 3E**). Thus, despite the lack of perfect subspace orthogonality, a downstream neural population would still be able to leverage the different subspaces to successfully bind both offer value to location and offer value to position in almost every case. Further, the theoretical result that the binding error rate depends only on the nonlinear distance implies that low binding error rates can be achieved even for subspace correlations that are close to one (**Figure 3B**).

Finally, we also use the estimated distances to produce a prediction for the generalization error rate of both the value-space and value-time codes. Due to the significant linear components found for every brain region except PCC in the spatial code and pgACC in both codes, as well as the moderate nonlinear distances, these predicted generalization error rates are also mostly below chance (**Figure 3F**, open circles). However, in this case, we can also compute the empirical generalization error rate of a linear classifier – that is, a classifier trained to decode the value of left offers then tested only on right offers. This empirical generalization performance is nearly identical to the predicted performance (**Figure 3F**, outlined circles, and see **Figure S2** for the standard decoding performance). This agreement indicates that our theory captures the aspects of the representational geometry that are relevant to decoding performance. As a consequence, we believe it lends indirect support to the validity of the low misbinding error rate also predicted by our theory (**Figure 3E**). Together, this analysis framework identifies each of the regions that we studied here on the binding and generalization error rate plane (**Figure 3G**). Interestingly, most regions do not specialize toward either highly nonlinear or highly linear representations (thereby minimizing only one type of error); instead, they exist at a midpoint, where they simultaneously have large linear and nonlinear distances – and thereby reducing both forms of errors.

**Neurons with heterogeneous representations of value across positions drive subspace separation**

What is the single neuron basis of the subspaces? One simple possibility is that the subspaces may be composed almost entirely of nonlinearly selective neurons that each only represent the value of offers at a single spatial position (i.e., neurons with canonical spatial receptive fields). The full population, then, would be composed of two largely separate subpopulations: one for offers on the left and another for offers on the right (**Figure 4A**, left, and **4B**, top). This organization could also be viewed as a gain modulation code, where spatial position strongly modulates the value code of single neurons, without changing their tuning (e.g., **Figure 4E**, left: a left offer tuned cell). Alternatively, subspace separation could be achieved by neurons with heterogeneous nonlinear responses to offer value and position, and that contribute activity to multiple subspaces (e.g., **Figure 4E**, right; Fusi et al., 2016; Tang et al., 2020). We refer to this as the *shared population hypothesis* (**Figure 4A**, right, and **4B, bottom**).

These different coding strategies lead to distinct predictions across the population. If the population was dominated by cells that respond more strongly to offers presented at one of the two positions (or even cells that respond only to offers presented at one position), then we would expect a bimodal distribution of differences in selectivity for left and right value across the population (**Figure 4C;** Elsayed et al., 2016). Alternatively, if the population is composed of neurons that represent value differently across the two positions, but with similar strength, then we would expect this distribution to be unimodal (**Figure 4C**). We discriminated between these two possibilities using the regression coefficients to characterize each neuron's value representation for each location. We then asked if the distributions of firing rate differences diverged from a unimodal distribution using Hartigan's dip test. We then repeated this analysis for all time-windows. We found no evidence for bimodality ($p > 0.9$ for all areas and subjects). The first offer on time window provides an example of the typical distribution (**Figure 4D**). This result therefore supports the idea that the value of left and right offers are encoded in distinct subspaces, but not in distinct populations of neurons (as would be expected by a simple spatial receptive field model). Note that this finding of non-categoricality is consistent with several other recent studies emphasizing non-categorical neural responses (Blanchard et al., 2018; Raposo et al., 2014; Kaufman et al., 2022).

The lack of bimodality implies there may be several types of nonlinear encoding neurons that drive the subspaces. Next, we show the left and right subspaces are supported by neurons with both spatially-tuned gain modulation and heterogenous value representations (e.g., **Figure 4E**). Specifically, we quantified the proportion of each neuron type in each brain region. First, we searched for neurons that contributed to both a left and right subspace within an analysis epoch, using a measure of subspace contribution (Xie et al., 2022; **Methods**). The subspace contribution effectively computes the proportion of variance a neuron contributes to a subspace. Next, remaining neurons were classified based on whether their preferred value was either constant (gain modulated) or shifted across subspaces (heterogeneous nonlinear). Examining the response profiles of neurons meeting the multi-subspace criterion shows both those with gain modulation and those with shifting tuning to value that depends on the subspace (**Figure 4E**). Averaging across all regions and windows, we found that both types were found in roughly equal proportions. These results rule out simple spatial subpopulation or wide-spread gain modulation explanations of value subspaces and support the idea that the two subspaces are represented in the same population of neurons, but semi-orthogonal axes in population space.

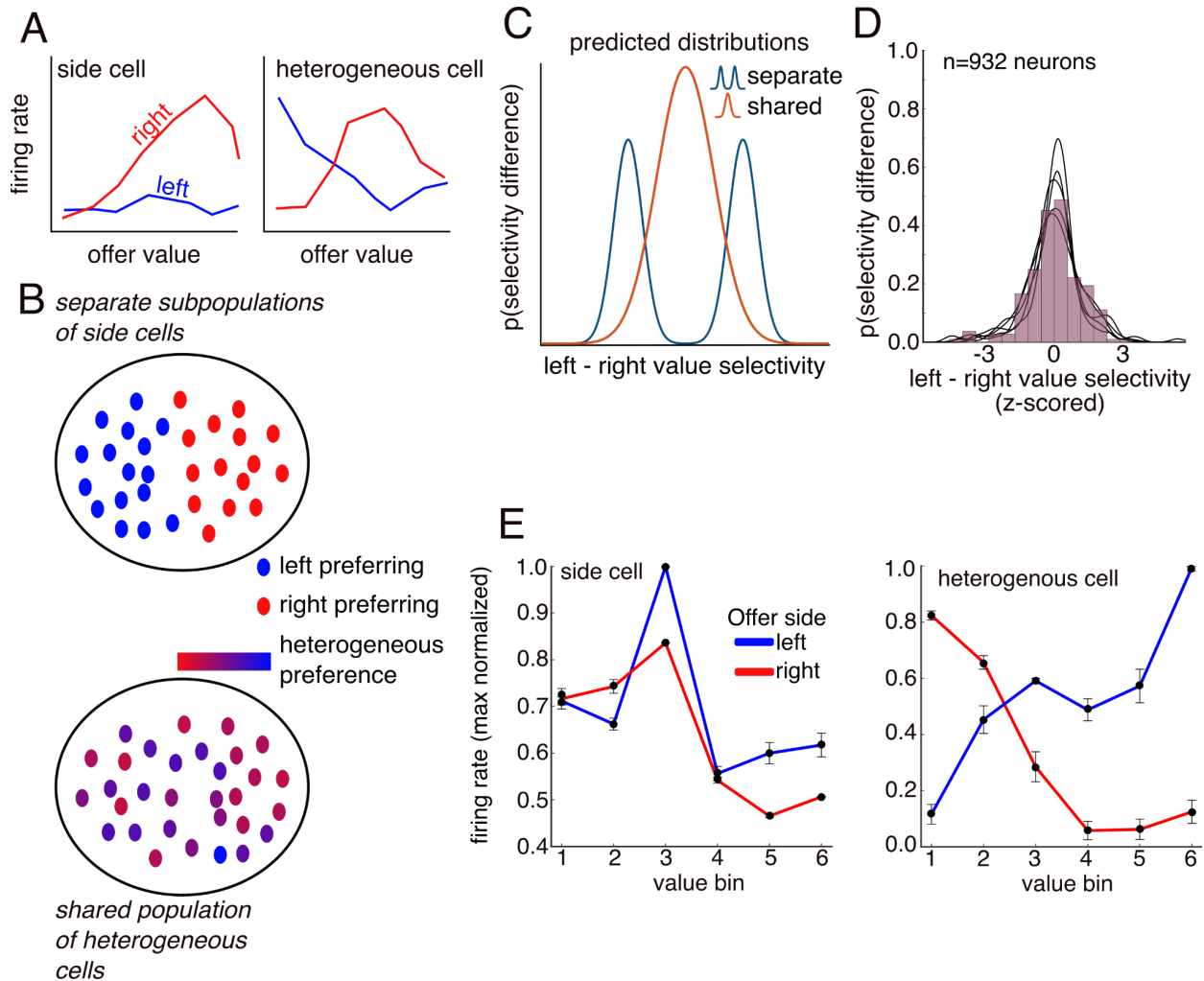

**Figure 4.** Understanding the nonlinear selectivity underlying subspaces and binding. **A.** Two schematic kinds of nonlinear selectivity: (left) A strong side preference and (right) two different response profiles to the two different sides. **B.** The two kinds of selectivity give rise to two distinct hypotheses for selectivity across the population: (top) Separate subpopulations, each composed of cells with a strong preference for one of the two positions and (bottom) a single population composed of cells with heterogeneous response profiles for the two sides. Both forms of nonlinear population selectivity achieve subspace binding. (**C**). These hypotheses (**A-B**) predict differences in how the distribution of selectivity differences for left and right value subspaces will appear. The separate subpopulations hypothesis predicts a closer to bimodal distribution (blue line),

while the shared, heterogeneous hypothesis predicts a unimodal distribution (orange line). (**D**) estimated distribution of differences in value selectivity for left and right subspaces for offer 1 on time window. Each line is a different region, showing they are all unimodal. (**E**). Example OFC nonlinear encoding neurons showing: one has a weak side preference (left) and the other has a heterogeneous response profile (right).

# DISCUSSION

We asked how value is bound to space in neural populations from five core cortical reward areas. Our central hypothesis, which our data support, is that in a binary choice task, the value of left and right offers are encoded in semi-orthogonal subspaces. This encoding scheme supports both reliable binding of value to offer position, and reliable generalization of the value code across both positions. In other words, this orthogonalization was sufficient to allow binding of information about value to information about space, thereby facilitating the selection of the appropriate action. Indeed, we find that session-to-session variability in the strength of orthogonalization correlates with choice accuracy. This result supports a link between subspace orthogonalization behavior. Together, these results suggest a novel solution to a classic problem in neuroeconomics, one that is grounded in the remarkable coding properties of neuronal populations. Moreover, they raise the possibility that subspace orthogonalization may be a general solution to other important binding problems, such as the perceptual binding problem (Treisman & Gelade, 1980).

Why are subspaces semi-orthogonal, rather than fully orthogonal? Both orthogonal and semi-orthogonal subspaces allow for the binding of value to space. However, the semi-orthogonal subspaces that we find here are simultaneously collinear enough to permit generalization of value across offer positions (Bernardi et al., 2020). A code that permits generalization of value across spatial positions has several benefits for learning and generalizing the mapping from visual stimuli to value: First, it would allow the mapping to be learned simultaneously for multiple different offer positions, thus allowing learning from fewer trials than if the mapping had to be learned separately for every location. Second, it would permit the animal to generalize the mapping it already knows to novel offer positions, thus allowing for significant transfer to novel situations. Further, unlike previous theoretical work (Barak, 2013), this decomposition formulates a link from neuron selectivity to generalization that is outside of the training distribution underlying the decoder. The generalizability of value is also in line with results from other studies examining perceptual learning of magnitude and ordinal rank, which show generalizability of other scalar parameters (Sheahan et al., 2021). Indeed, such scales support the general utility of rapidly mapping stimuli to novel contexts. Together, the semi-orthogonal subspaces composed of their relative linear and nonlinear power suggest the neural geometry is in a sweet spot that balances generalization and binding.

The result that both low misbinding error rates and above-chance generalization can co-exist is interesting given debates about the relative benefits of low- versus high-dimensional neural geometries (Cueva et al., 2020; Flesch et al., 2022; Gallego et al., 2017; Jazayeri and Ostojic, 2021). On one hand, some previous findings have pointed to nonlinearly mixed representations of different latent variables with high embedding dimensionality, particularly in prefrontal areas, that can support strong binding and provide flexibility for a linear readout (Bernardi et al., 2020; Rigotti et al., 2013). In contrast, others have pointed to how factorized representations of different latent variables (often with low embedding dimensionality) can support generalization across contexts, rapid learning, and reliably encoding (Bernardi et al., 2020; Cueva et al., 2020; Flesch et al., 2022; Gallego et al., 2017; Sohn et al., 2019). These differences have driven debates about how dimensionality should be interpreted with respect to linking neural coding and behavior (Jazayeri and Ostojic, 2021). However, deeper interrogation suggests that this is a false dichotomy. In particular, we show that a mixture of low and high dimensional representations can simultaneously achieve the benefits of both forms of representation – that is, there can exist both a low-dimensional scaffold to ensure good generalization, and enough high-dimensional nonlinear mixing to ensure reliable downstream read-out of bound features, consistent with other work in mice and monkeys that show a similar intermediate geometry (Bernardi et al., 2020; Nogueira et al., 2021; Boyle et al., 2022). Thus, the single measure of embedding dimensionality falls short of a full description of the properties of the corresponding code. We believe that deeper progress in connecting the geometry of neural populations and networks to behavior can be made by investigating the strength of these different representational components separately, and quantifying what this means for the downstream ability to decode, bind, and generalize based on these representations. Such ideas about the structure of networks and population code properties underlying binding, generalization and compositionality have begun to help explain constraints in the capacity for cognitive control (Musslick and Cohen, 2021).

While we have focused on representations of offer value that are tied to specific spatial positions (i.e., offers presented on the left and right), this representation can coexist with representations that depend on or are more closely tied to the animal's eventual choice. In particular, some previous studies have found increased representation of the eventually chosen offer (Strait et al., 2014) or representations of the difference in value between the two offers, rather than the value of each offer separately (Strait et al., 2015). The existence of these other representations alongside the spatially bound representations that we study here would not change the main conclusions of our study. Further, the spatially bound representations that we study may underlie the formation of these other representations. For instance, a spatially bound representation of the first offer value would be useful for developing a representation of the difference in offer values across the two spatial positions. From an analysis perspective, not including terms for chosen and unchosen offers makes our model less specified rather than incorrectly specified. While this would bias our results given strong choice encodings, the bias would manifest equally in both the measured subspace correlation value and the noise threshold – thus, the qualitative results reported here would not change. Taking our results alongside the conclusions from previous studies suggests that neurons in core value regions encode offer value bound to spatial position, along with a nonlinear modulation by the animal's choice.

We and others have noted the encoding of spatial information in single neurons in value-sensitive regions of the brain (Strait et al., 2016; Yoo et al., 2018; Roesch and Schoenbaum, 2006; Feierstein et al., 2006; Tsujimoto et al., 2009; Luk and Wallis, 2013). However, the functional meaning of such signals has been unclear. We have previously conjectured that these spatial signals modulate the encoding of task variables like value through gain-like changes that do not alter tuning (Strait et al., 2016; Yoo et al., 2018; Hayden and Moreno-Bote, 2018). Our present results indicate that our previous hypotheses were partially incorrect. Instead, it also appears that space alters the tuning profile of value-sensitive neurons. In other words, space alters firing, but in many cases, a neuron exhibits nonlinear, heterogeneous tuning to space and value; their preferred value changes in a subspace dependent manner. Neurons that exhibit a shift in preferred tuning are expected when they interact with gain modulation neurons in networks with diverse and recurrent connectivity (Salinas and Sejnowski, 2001; Zhang and Abbott, 2000).

If gain modulated neurons do not provide the full solution to the binding problem, then what is the computational process in the network that supports binding value to space? Standard approaches tend to start with the idea that networks solve this problem through specialized subpopulations, such as gain modulators (Botvinick and Watanabe, 2007; Pouget and Sejnowski, 1997; Salinas and Sejnowski, 2001). However, our current results point to an amalgamation of heterogeneous and gain modulated neurons, requiring a computation that is likely rendered through a tensor-like operation through the network's synaptic weights. This population-level operation predicts value should be clearly demixed in our populations. Indeed, we find a canonical ordering of value when considering the projection of single neuron codes into a population read-out of value (**Figure S3A**). The value- and value-side specific subspace code suggests space acts as a gain modulator (**Figure S3B-E**) at the population level to solve the binding problem and provide a demixed representation of value according to the offer side; this idea is in line with similar findings in sequence working memory (Xie et al., 2022).

The question of how decision-related information such as value is transformed into action is one of the major ones in the field of neuroeconomics (Kable and Glimcher, 2007; Krabijch et al., 2010; Knudsen and Wallis, 2022; Rangel et al., 2008; Padoa-Schioppa and Assad, 2006). Typical theories hold that values are represented in an abstract value space that is conceptually and functionally distinct from the action space needed to implement the choice of the action. As such, there is a question of how value links up with action. This problem is often reified in neuroanatomy - some regions are assumed to be pure value regions while other, presumably anatomically downstream regions, are assumed to have action signals (Rangel et al., 2008 Kable and Glimcher, 2009; Padoa-Schioppa, 2011). Our results here suggest a somewhat different conclusion. They suggest that the same neural code provides an abstract representation of value as well as representations of value that are bound to particular spatial positions. Further, we do not find that the neurons supporting these different aspects (abstract as opposed to spatially bound) are divided into distinct subpopulations. This argues against the idea that there is an anatomical distinction between value and action frames (Fine and Hayden, 2022; Hayden and

Niv, 2021). Our findings suggest that this mechanism for binding through semi-orthogonal

subspaces is used throughout the cortical reward system. Together, these results highlight the

potential value of functional specialization through population representation, rather than

through modular architecture, for solving long-standing problems in neuroeconomics (Ebitz and

Hayden, 2021; Urai et al., 2021; Saxena and Cunningham, 2019).

# METHODS

*Surgical procedures.* All procedures were approved by either the University Committee on Animal Resources at the University of Rochester or the IACUC at the University of Minnesota. Animal procedures were also designed and conducted in compliance with the Public Health Service's Guide for the Care and Use of Animals. All surgery was performed under anesthesia. Male rhesus macaques (*Macaca mulatta*) served as subjects. A small prosthesis was used to maintain stability. Animals were habituated to laboratory conditions and then trained to perform oculomotor tasks for liquid rewards. We placed a Cilux recording chamber (Crist Instruments) over the area of interest. We verified positioning by magnetic resonance imaging with the aid of a Brainsight system (Rogue Research). Animals received appropriate analgesics and antibiotics after all procedures. Throughout both behavioral and physiological recording sessions, we kept the chamber clean with regular antibiotic washes and sealed them with sterile caps.

*Recording sites.* We approached our brain regions through standard recording grids (Crist Instruments) guided by a micromanipulator (NAN Instruments). All recording sites were selected based on the boundaries given in the Paxinos atlas (Paxinos et al., 2008). In all cases we sampled evenly across the regions. Neuronal recordings in OFC were collected from subjects P and S; recordings in rOFC were collected from subjects V and P; recordings in vmPFC were collected from subjects B and H; recordings in pgACC were collected from subject B and V; recordings from PCC were collected from subject P and S; and recording in VS were collected from subject B and C.

We defined **OFC 11/13** as lying within the coronal planes situated between 28.65 and 42.15 mm rostral to the interaural plane, the horizontal planes situated between 3 and 9.5 mm from the brain's ventral surface, and the sagittal planes between 3 and 14 mm from the medial wall. The coordinates correspond to both areas 11 and 13 in Paxinos et al. (2008). We used the same criteria in a different dataset (Blanchard et al., 2015).

We defined **vmPFC 14** as lying within the coronal planes situated between 29 and 44 mm rostral to the interaural plane, the horizontal planes situated between 0 and 9 mm from the brain's ventral surface, and the sagittal planes between 0 and 8 mm from the medial wall. These coordinates correspond to area 14m in Paxinos et al. (2008). This dataset was used in Strait et al., 2014 and 2016.

We defined **pgACC 32** as lying within the coronal planes situated between 30.90 and 40.10 mm rostral to the interaural plane, the horizontal planes situated between 7.30 and 15.50 mm from the brain's dorsal surface, and the sagittal planes between 0 and 4.5 mm from the medial wall. Our recordings were made from central regions within these zones, which correspond to area 32 in Paxinos et al. (2008). Note that the term area 32 is sometimes used more broadly than we use it here, and in those studies encompasses the dorsal anterior cingulate cortex; we believe that that region, which is not studied here, should be called area 24 (Heilbronner and Hayden, 2016).

We defined **PCC 29/31** as lying within the coronal planes situated between 2.88 mm caudal and 15.6 mm rostral to the interaural plane, the horizontal planes situated between 16.5 and 22.5 mm from the brain's dorsal surface, and the sagittal planes between 0 and 6 mm from the medial wall. The coordinates correspond to area 29/31 in Paxinos et al. (2008, Wang et al., 2020).

We defined **VS** as lying within the coronal planes situated between 20.66 and 28.02 mm rostral to the interaural plane, the horizontal planes situated between 0 and 8.01 mm from the ventral surface of the striatum, and the sagittal planes between 0 and 8.69 mm from the medial wall. Note that our recording sites were targeted towards the nucleus accumbens core region of the VS. This dataset was used in Strait et al. (2015 and 2016).

We confirmed the recording location before each recording session using our Brainsight system with structural magnetic resonance images taken before the experiment. Neuroimaging was performed at the Rochester Center for Brain Imaging on a Siemens 3T MAGNETOM Trio Tim using 0.5 mm voxels or in the Center for Magnetic Resonance Research at UMN. We confirmed recording locations by listening for characteristic sounds of white and gray matter during recording, which in all cases matched the loci indicated by the Brainsight system.

*Electrophysiological techniques and processing.* Either single (FHC) or multi-contact electrodes (V-Probe, Plexon) were lowered using a microdrive (NAN Instruments) until waveforms between one and three neuron(s) were isolated. Individual action potentials were isolated on a Plexon system (Plexon, Dallas, TX) or Ripple Neuro (Salt Lake City, UT). Neurons were selected for study solely on the basis of the quality of isolation; we never preselected based on task-related response properties. All collected neurons for which we managed to obtain at least 300 trials were analyzed; no neurons that surpassed our isolation criteria were excluded from analysis.

*Eye-tracking and reward delivery.* Eye position was sampled at 1,000 Hz by an infrared eye-monitoring camera system (SR Research). Stimuli were controlled by a computer running Matlab (Mathworks) with Psychtoolbox and Eyelink Toolbox. Visual stimuli were colored rectangles on a computer monitor placed 57 cm from the animal and centered on its eyes. A standard solenoid valve controlled the duration of juice delivery. Solenoid calibration was performed daily.

*Risky choice task.* The task made use of vertical rectangles indicating reward amount and probability. We have shown in a variety of contexts that this method provides reliable communication of abstract concepts such as reward, probability, delay, and rule to monkeys (e.g. Azab et al., 2017 and 2018; Sleezer et al., 2016; Blanchard et al., 2014). The task presented two offers on each trial. A rectangle 300 pixels tall and 80 pixels wide represented each offer (11.35° of visual angle tall and 4.08° of visual angle wide). Two parameters defined gamble offers, *stakes* and *probability*. Each gamble rectangle was divided into two portions, one red and the other either gray, blue, or green. The size of the color portions signified the probability of winning a small (125 μl, gray), medium (165 μl, blue), or large reward (240 μl, green), respectively. We used a uniform distribution between 0 and 100% for probabilities. The size of the red portion indicated the probability of no reward. Offer types were selected at random with a 43.75% probability of blue (medium magnitude) gamble, a 43.75% probability of green (high magnitude) gambles, and a 12.5% probability of gray options (safe offers). All safe offers were excluded from the analyses described here, although we confirmed that the results are the same if these trials are included. Previous training history for these subjects included several saccade-based laboratory tasks, including a cognitive control task (Hayden et al., 2010), two stochastic choice tasks (Blanchard et al., 2014; Heilbronner and Hayden, 2016), a foraging task (Blanchard and Hayden, 2015), and a discounting task (Pearson et al., 2010).

On each trial, one offer appeared on the left side of the screen and the other appeared on the right. We randomized the sides of the first and second offer. Both offers appeared for 400 ms and were followed by a 600-ms blank period. After the offers were presented separately, a central fixation spot appeared, and the monkey fixated on it for 100 ms. Next, both offers appeared simultaneously and the animal indicated its choice by shifting gaze to its preferred offer and maintaining fixation on it for 200 ms. Failure to maintain gaze for 200 ms did not lead to the end of the trial but instead returned the monkey to a choice state; thus, monkeys were free to change their mind if they did so within 200 ms (although in our observations, they seldom did so). Following a successful 200-ms fixation, the gamble was resolved and the reward was delivered. We defined trials that took > 7 sec as inattentive trials and we did not include them in the analyses (this removed ~1% of trials). Outcomes that yielded rewards were accompanied by a visual cue: a white circle in the center of the chosen offer. All trials were followed by an 800-ms intertrial interval with a blank screen.

*Behavioral control for neural data indicates decisions from online valuation comparison.* The aim in this task design is to examine the process of valuation and decision-making in a sequential setting. In our design, the subjects are presented with the offers again to allow them the chance to reify their choice. To ensure that their choices are actually based on the online, sequential presentation of the offers in the main task, we examined previous behavioral data (in 7 sessions for each of 4 subjects) collected in a version of this task where the 2nd presentation does not occur, and they have to respond from memory. These results are also reported elsewhere (Yoo and Hayden, 2019). The result of these data was that subjects still chose the higher value offer approximately ~80% of the time. This result aligns with the findings with the main task data presented here, wherein (all) subjects chose the higher value approximately %82 of the time.

*Behavioral analysis, model estimation and subjective value.* The decision-variables underlying subject choice could arise from several possible estimates over probability, stakes and the estimated value. Our previous analysis and modeling of the present behavioral data indicate that the monkeys in this task make choices in line with a subjective valuation estimate of offers that reflects subjective attitudes towards the offer size (stakes) that indicate risk-seeking, and a warped probability estimation fit well by a prelec warping function. We remodeled the decisions here to derive subjective value estimates for using in estimation of neural encoding subspaces (see below) and establishing the connection of these subspaces to monkey choices. Here, we consider models where the monkey's subjective value (SV) for an offer follows the probability times the stakes.

The model space we considered included 4 models. The models included those where the stakes were either assumed to be observed objectively or weighted as a power-law, and the probability was either observed objectively or transformed with a prelec function. For all models, choice probabilities were assumed to be generated by a softmax decision function over the relative subjective value. For example, during model fitting of a model with subjective utility (weighted stakes) and weighted probability, 3 model terms were fit. The terms for utility and probability, and the softmax temperature term. All models were fitted using the variational bayesian toolbox, and model comparisons performed using bayesian model selection over the model free energies (Daunizeau, Adam, & Rigoux, 2014).

**Neural processing, data-selection, and statistical analysis**
We calculated firing rates in 20-ms bins but we analyzed them in longer (400 ms) epochs. For estimating regression coefficients (see below), firing rates were all Z-scored for each neuron. In regression estimates, we only included risky trials. No smoothing was applied before any reported analysis.

*Single-Neuron linear regression model.* Individual neuron selectivity was estimated using a linear regression model. Model coefficients were estimated using the z-scored, time-averaged firing rate in 4 different task windows. The windows were 400 ms blocks, with two separate blocks in the offer 1 window and 2 more blocks in the offer 2 window. The first blocks in a window were from 0 (offer on) to 400 ms after (offer off), and the second window was from 450 to 850 ms after the offer was presented. We refer to the latter window as the delay window. The estimated model had coefficients for subjective value, offer spatial position, and their interaction which can be used to estimate the nonlinear encoding terms. Model terms for offer side were effects coded [-1 1]; expected value was estimated with linear and b-spline terms (see **Figure S1** for more details on the splines), with the value term being min-max normalized to standardize across units and monkeys. The design matrix in total therefore had 5 variables to estimate plus the intercept. The model formula for a neuron (n) firing rate (FR) was estimated using all trials:

$$\text{FR(n, trials)} = \beta_0 + \beta_1 \text{Value} + \beta_2 \text{Space} + \beta(\text{Interaction}) + N(0, \sigma^2)$$

*Computing population encoding subspaces for space and value.* To measure separability between spatially distinct value subspaces, we used the computed regression coefficients for calculating the population encoding subspaces. Our goal was to compare the value subspaces for left and right offers both within offer epochs (e.g., offer 1) and across offer epochs. Because the coefficients ($\beta$) were derived from a linear model, we added up the coefficients to create a value vector for each distinct side and epoch.

Computation of each subspace involves setting the levels of X in the $\beta X$ of the regression equation, given an already set of fit $\beta$s per neuron (see *Single-Neuron linear regression model* above). Essentially, this process gives the model predicted firing rate spanning from the lowest range normalized value offer that includes the intercept offset (0) to the highest (1). The process of creating the value subspace vector for the left-side, for example, proceeded as follows: first, we must subtract two vectors, with the base vector centered on the different sides being compared (e.g., left and right), as they themselves have different intercepts. Given the linear model form for the design matrix, and assuming left == 1 and right == -1, the intercept vector for value ==0 and left offer subspace, the vector is:

$$\text{Value high vector: } [1]\beta_0 + [1]\beta_{\text{Value}} + [1]\beta_{\text{Space}} + [1]\beta_{\text{Interaction}}$$
$$-$$
$$\text{Value/Side intercept vector: } [1]\beta_0 + [0]\beta_{\text{Value}} + [1]\beta_{\text{Space}} + [0]\beta_{\text{Interaction}}$$

In total, there were six distinct task-functional comparisons, including comparisons of (1) left and right within offers 1 and 2, (2) left and right between offers 1 and 2, and (3) between the same-side across offers (i.e., left and left for offer 1 and 2, right and right for offer 1 and 2).

The testing of separability (or orthogonality) between subspaces requires establishing a null hypothesis of 1. This is because separable subspaces will be less correlated than a perfect correlation under the influence of noise. Specifically, because our main hypothesis was essentially a close to zero correlation between subspaces (i.e., |r|=0), we needed to estimate how a perfect correlation in the dataset, but confounded by noise, would be distributed (|r|=1). Therefore, we consider a subspace as separable or (semi-)orthogonal if the correlation between neuron weights is outside the confidence bound of this hypothetical perfect correlation distribution.

We addressed this problem by applying an already established approach (Kimmel et al., 2020). In brief, we fit the linear models described above on 1000 bootstrap resampled sets of trials. For each set, we computed the subspace correlation between left and right value representations. Then, to construct the noise ceiling estimate we computed the subspace correlation between the left value subspaces for the first and last 500 resamples (yielding 500 subspace correlation estimates) as well as the correlation between the right value subspaces from the first and last 500 resamples (yielding another 500 estimates). We then compared the 1000 subspace correlation estimates between the left and right value subspaces with the 1000 total estimates from the noise ceiling distribution using the bootstrap test.

Finally, we repeated this same procedure for the b-spline value representation models, using the alignment index (Kimmel et al., 2020) instead of the subspace correlation. These results are shown in **Figure S1**.

*Comparison between models with parallel and non-parallel subspaces.* To understand whether the single neuron basis of our semi-orthogonal subspaces could be explained equally well as noise, we performed a model comparison analysis for three classes of models: First, a regression model that fit only a noise term, and therefore explained the data only as random fluctuations. Second, models that had both side and value terms, but with no interaction between them. At the population level, this model would produce parallel subspaces. We fit these models with both linear and nonlinear representations of value (see **Figure S1** for detail). Third, models that have both side and value terms as well as an interaction term between them. If the interaction term is significant, then, at the population level, this model produces non-parallel subspaces.

For each model, we obtained a posterior predictive distribution, which is a distribution of samples from the model combined across all trials. We can use this posterior predictive distribution to check the fit of our model, and we find that our models provide good fit to the value response curves of the actual neurons (**Figures 2B** and **S3B**). Then, we perform a Bayesian model stacking analysis which assigns a weight to each model (Yao et al., 2017). In the stacking analysis, the posterior predictive distributions of all the models are combined to produce a combined model (i.e., a weighted sum of posterior predictive samples) that maximizes the leave-one-out cross-validation performance. We obtain similar results for more traditional measures of model goodness of fit, such as the widely applicable information criterion (Watanabe, 2013). However, we believe the stacking analysis provides useful intuition: It both gives insight into which model provides the best fit, as well as which combination of models provides the best fit.

*Linking Subspace separability to behavior.* To determine whether the extent of subspace separability was related to later choice behavior, we conducted a bootstrap process of fitting behavioral models, linear regression models, and computing subspace correlations. Specifically, we hypothesized that if the subspace correlation was indicative of binding and offer partitioning between offer 1 and 2, then more separable subspaces should correlate with fewer suboptimal choices.

To test this, we needed populations of neurons that were collected simultaneously with choice. To determine the minimum number of neurons needed for analysis, we approximated an effective minimum sample size. Under the assumption of normally distributed regression coefficients across neurons, we asked at what number of neurons the variance of the distribution of regression coefficients became stable. In other words, for each monkey, we drew random subpopulations of neurons and computed the variance of the coefficients within that subpopulation. We drew populations from size 5 to 55, and drew 500 random subsamples for each population size. We chose the effective sample size as the point at which this variance became unchanging, reflecting an approximately stable estimate of the distribution of coefficients. The sample size estimate averaged to 28 neurons across monkeys, and became our minimum for determining eligible monkeys.

Because our aim was to determine whether a positive correlation existed between subspace correlation and # of suboptimal choices, we created 500 bootstrap estimates of the subspace correlation for each monkey separately in each session with simultaneous neurons. Within each bootstrap, we also computed the # of suboptimal choices as the percent of times choosing the lower subjective value. To test for significance and control for chance correlations, we created a null distribution of correlations by randomly permuting the subspace correlations and then correlating those randomized vectors with the # of suboptimal choices. We computed the final p-value of the empirical correlation between subspace correlation and suboptimal choice for each monkey and session as a z-score:

$$p = 1 - normcdf\left(\frac{\rho_{empirical} - \mu\left(\rho_{permuted}\right)}{\sigma\left(\rho_{permuted}\right)}\right)$$

*Neuron selectivity.* We determined the single neuron selectivity using a permutation ANOVA. For each neuron, we first fit an ANOVA using the same model terms as for the linear regression above. The only difference between the regression and ANOVA approach was that the ANOVA treated value as categorical and discretized over 7 equally distributed levels of value per monkey and per session. To control for chance detection of effects, we obtained a null distribution of ANOVA F-scores by re-running the ANOVA 1000 times per neuron using random permutations of the firing rate. The p-value was computed by for each effect and interaction as:

$$p = \frac{\#(Fperm > F_{main}) + 1}{\#(Fperm) + 1}$$

We used these p-values to determine whether a neuron was linearly selective for a single variable, linear mixed selectivity (both value and side terms significant), or nonlinearly selective (significant interaction term). This procedure was repeated across all 4 time windows, including both offer on epochs and the both offer delay windows. The proportion of each type of selectivity was determined using all monkeys within a brain region.

*Single-Neuron basis of subspaces: bimodality.* To test whether the subspaces are primarily formed from gain modulating neurons, we note that the distribution of firing rate differences (across space) to value tuning should deviate from unimodal. We tested this by asking whether the distribution of differences in value tuning vectors for left and right offers was bimodal, using Hartigan's dip test of bimodality. Specifically, we formed distributions of tuning differences using the left and right value subspace coefficients for each neuron computed above.

The distributional responses were computed for each offer (1 and 2) and time-window (offer on and delay) as a sensitivity index:

$$\text{Value-space firing rate (FR) sensitivity} = \frac{FR(n)_{left} - \mu(FR(n)_{left})}{\sigma(FR(n)_{left})} - \frac{FR(n)_{right} - \mu(FR(n)_{right})}{\sigma(FR(n)_{right})}$$

In the above, for example, $FR(n)_{left}$ is the value-left subspace vector coefficient for neuron $n$.

*Single-Neuron basis of subspaces: subspace contribution of gain modulation versus complex nonlinear.* The bimodality test described above provides a global measure of whether the populations of neurons composing the subspaces are generally specialized to certain spatial locations. To further demarcate the contribution of gain versus complex nonlinear encodings to the subspaces, we use the property that gain modulated neurons will have the same preferred response to value, but with different amplitudes for different spatial positions, while complex nonlinear neurons will have a shifting preference to value. The main result we aimed for was determining the prevalence of nonlinear complex versus gain modulating neurons, and determining their % contribution of each type in forming the subspaces. To do this, we first determined which neurons in a region contributed to both left and right value subspaces for each epoch and time-window. Subspace contribution was determined by finding each neuron's percent variance in each subspace:

$$\text{Subspace contribution (\%)} = \frac{(FR(n)_{space}^2)^2}{\sum (FR(n)_{space}^4)}$$

The above neuron participation ratio is inherently related to the dimensionality of that subspace; effectively, each neurons % variance contributed can be viewed as an approximation to the eigenvalues of the subspace (Gao et al., 2017; Xie et al., 2022). We retained the top 95% of neurons, which was an average of 48 neurons in each subspace, across all brain regions. We then compared whether each neuron's preferred value was the same in left and right subspaces, within each of the analysis windows used for the regressions. Specifically, we performed a bootstrap hypothesis test of mean differences. The mean firing rates for testing were re-sampled 1000 times for each neuron, and the mean rate was computed over 7 equally distributed levels of value, separately for left and right offers. The peak rate for each resample was collected in a vector and subsequently randomized 1000 times (with resampling) to create a null distribution of mean differences in firing preference. The p-value was computed by counting the # of times the randomly permuted mean difference was larger than the empirically estimated.

**Binding by subspace orthogonalization**

We begin by considering a form of neural code that does not reliably bind the different features of a single stimulus together. In particular, a factorized representation of different stimulus features – even if the code for each individual feature is nonlinear – will fail to distinguish between different possible stimulus set bindings. In our setting, the neural population responds to both offer position and offer value. On a single trial the animal is shown a set of two stimuli, $X$, which each have a position and value. So, this set can be written as $X = \{[p_1, v_1], [p_2, v_2]\}$, where $p_1$ is the position of offer 1 and $v_1$ is its value. Then, if the average response of a neural population $\bar{r}$ is given by a function $f(X) = \sum_x^X f(x)$, we can write

$$\bar{r} = f([p_1, v_1]) + f([p_2, v_2])$$

Now, if the function $f$ can be factorized into a sum of functions, $f_{pos}$ for offer position and $f_{val}$ for offer value, then we can rewrite the response as

$$\bar{r} = f_{val}(v_1) + f_{pos}(p_1) + f_{val}(v_2) + f_{pos}(p_2)$$

This response contains all the original information about the features of the two offers – but it does not preserve their binding. In particular, while a maximum likelihood decoder would have a maximum at the correct stimulus set $X$, it would also have an equivalent maximum at the chimeric stimulus set $S = \{[p_1, v_2], [p_2, v_1]\}$. This is because the average population response to the two stimulus sets is exactly the same,

$$
\begin{aligned}
\Delta &= f(X) - f(S) \\
&= f_{val}(v_1) + f_{pos}(p_1) + f_{val}(v_2) + f_{pos}(p_2) \\
&\quad -f_{val}(v_2) - f_{pos}(p_1) - f_{val}(v_1) - f_{pos}(p_2) \\
&= 0
\end{aligned}
$$

Thus, in the case of a high-value offer on the left and a low-value offer on the right, an alternative and equally likely interpretation of the neural representation would be that there was a low-value offer on the left and a high-value offer on the right, which would lead to a suboptimal choice by the animal.

In practice, humans and other animals may make these kinds of binding errors, but they are thought to be infrequent (Bays et al., 2022). So, in many cases, this ambiguity must be successfully resolved. To understand how this happens, we consider the case when $f$ cannot be fully factorized and instead can be written as $f([p, v]) = f_{pos}(p) + f_{val}(v) + f_{pos-val}([p, v])$, where $f_{pos-val}$ is non-factorizable function of both side and value. We refer to this term as the conjunctive part of the response. Then, the difference in response for the correct and chimeric stimulus sets $\Delta$ from before can be written as,

$$
\begin{aligned}
\Delta &= f(X) - f(S) \\
&= f_{pos-val}([v_1, p_1]) + f_{pos-val}([v_2, p_2]) - f_{pos-val}([v_2, p_1]) - f_{pos-val}([v_1, p_2])
\end{aligned}
$$

Thus, so long as the squared sum of $\Delta$ across the neural population is larger than zero – and, in a noisy system, larger than the noise – then a decoder will be able to resolve the ambiguity between the correct and chimeric stimulus set and avoid misbinding errors. We derive an expression for this misbinding rate in a simplified case below. One of our key theoretical results is that the inclusion of this conjunctive part of the representation can simultaneously resolve the coding ambiguity without destroying the benefits of a factorized representation (see X).

At the level of single neurons, the conjunctive part of the representation $f_{pos-val}$ manifests as neurons with nonlinear mixed selectivity for offer value and position. At the population level, the conjunctive part of the representation manifests as orthogonalization of the subspaces encoding the value of left and right offers.

**Linear-nonlinear code framework**

To apply our mathematical framework to the experimental data, we consider a discretized offer value along with offer position – value is discretizd as for our decoding analysis (see *in* ). In these data, we found that decoding a binary value yielded much better performance than decoding continuous value with various methods. The analytic approach we take below is also simplified for discrete value. However, the subspace binding hypothesis does not depend on this discreteness. Once we discretize value, we have $K = 2$ latent variables that each take on $n = 2$ different values. In this setting, we model the neural responses of $N$ neurons as,

$$r(x) = Lx_z + Mf_N(x) + \epsilon$$

where $r(x)$ is an $N \times 1$ vector of the z-scored activity of $N$ neurons, $L$ is an $N \times K$ linear transform of the z-scored stimulus vector $x_z$ and has columns $L = [d_{LV}L_1 \ d_{LA}L_2]$, $M$ is an $N \times n^K$ linear transform of $f_N(x)$, which is a nonlinear transform of the stimulus vector $x$ (defined in detail below). Finally, $\epsilon \sim \mathcal{N}(0, \sigma^2)$. A note: We z-score the stimulus vector $x$ (defined in detail below), so that the linear distance in the representation produced after the linear transform is controlled fully by the linear transform $L$, and does not depend on the specific choice of coding for $x$.

As mentioned above, each of the two features takes on two different values. So, $x$ is a vector with two elements that are each either 1 or 2. The first element corresponds to low- (1) or high-value (2); the second element corresponds to the two values of offer position. The possible $x$ are:

$$\begin{aligned} x_{11} &= [1 \ 1]^T \\ x_{12} &= [1 \ 2]^T \\ x_{21} &= [2 \ 1]^T \\ x_{22} &= [2 \ 2]^T \end{aligned}$$

The nonlinear function we consider is the conjunctive identity function used in (Rigotti et al., 2013; Johnston et al., 2020), where

$$f_N(x)_{ij} = [x_i = i][x_j = j]$$

and

$$f_N(x) = [f_N(x)_{11} \ f_N(x)_{12} \ f_N(x)_{21} \ f_N(x)_{22}]^T$$

For the linear transform $L$, which has $K = 2$ columns, the length of the first column is $d_{LV}$, which will be the average distance between two stimulus representations that differ only in their value (where $M = 0$). We will refer to the length of the second column as $d_{LA}$, the average distance between two stimuli that differ only in offer position (also where $M = 0$). All of the columns of $M$ will have the same length $m$, which will mean that the distance between the nonlinear components of two stimulus representations will be $d_N = \sqrt{2}m$. We assume that the length of the nonlinear perturbation for each stimulus is the same; while this is unlikely to be precisely true, it simplifies our analysis and still gives good results when comparing to the experimental data. We also assume that the columns of both $L$ and $M$ are orthogonal, both within each matrix and between the two matrices (though the analytic results are similar without the between matrix orthogonality). For large $N$, this will tend to be true for random vectors.

Finally, a code in this framework is described by a stimulus set ($K$ and $n$) and four parameters: $d_{LV}$ (representing offer value), $d_{LA}$ (representing offer position), $d_N$ (representing the nonlinear part of the code), and $\sigma$ (representing the standard deviation of the noise. In our analytic theory, we show that the binding error rate depends on $d_N$ and $\sigma$. That is, low binding error rates can be achieved so long as $d_N$ is sufficiently large, and does not depend on the linear part of the code ($d_{LV}$ or $d_{LA}$). We also show that the generalization error rate depends only on $d_{LV}$, $d_N$, and $\sigma$ – that is, it does not depend on $d_{LA}$. Thus, we will focus on estimating $d_{LV}$, $d_N$, and $\sigma$ from the experimental data.

**Linear-nonlinear data decomposition**

We want to estimate the three parameters of our code model, $d_{LV}$ (the linear code distance for value), $d_N$ (the nonlinear code distance), and $\sigma$ (the noise standard deviation), from the data.

To do this, we use a cross-validated distance measure, often referred to as the crossnobis distance (though we do not incorporate an estimate of the noise structure into our use of the measure, so it is not a version of the Mahalanobis distance as the name crossnobis suggests), to estimate the unbiased Euclidean distance between every pair of our four stimulus conditions. We used the routines provided by the python RSA toolbox (Nili et al., 2014). This yields a representational distance matrix, which can be decomposed according to our model framework. In particular, we minimize the reconstruction error between the estimated distance matrix and a distance matrix constructed only from $d_{LV}$, $d_{LA}$, and $d_N$ – as defined in the previous section. We find that this decomposition is stable from different initializations.

Next, we estimate the magnitude of the noise in the neural representations. Here, we choose to estimate it along a single dimension that is particularly relevant for our analyses. For our generalization analysis, $x_{11}$ and $x_{21}$ are defined as the training set while $x_{12}$ and $x_{22}$ are defined as the testing set. That is, the decoder will be trained to discriminate between $r(x_{11})$ and $r(x_{21})$ and then tested on its ability to discriminate between $r(x_{12})$ and $r(x_{22})$. Due to this, noise specifically along this learned decoding dimension is most relevant to generalization performance, and we will estimate specifically the magnitude of this noise from our data. In particular,

$$\sigma^2 = \mathbb{E}_{x_{ij}} \left[ \frac{v_1}{|v_1|_2} \cdot \left( r(x_{ij}) - \bar{r}(x_{ij}) \right) \right]^2$$

where the expectation is taken across both all trials from a particular condition and all conditions defined by $i$ and $j$.

Finally, we also compute the standard error of the mean of each of our estimates, collapsed into a standard error distance $\epsilon$, which will affect generalization performance, as described below.

**Relating subspace correlation to linear and nonlinear distance**

The linear-nonlinear code framework developed here provides a different way to define the subspace correlation measured used in the rest of the paper. We compute subspace correlation directly from the linear ($d_{LV}$) and nonlinear distances ($d_N$). In particular, we can take the cosine similarity between $v_1$ and $v_2$, which are defined in *[rep-vectors]*,

$$\rho = \frac{v_1 \cdot v_2}{|v_1|_2 |v_2|_2}$$
$$= \frac{(L_1 + M_1 - M_2) \cdot (L_1 + M_3 - M_4)}{d_{LV}^2 + d_N^2}$$
$$= \frac{d_{LV}^2}{d_{LV}^2 + d_N^2}$$

where $\rho$ is the subspace correlation (and see **Figure 3A** for a schematic).

**Derivation of the binding error rate**

We begin by considering a purely linear code for multiple stimuli $X$, where

$$r_L(X) = \sum_{x \in X} L\, x_z + \epsilon$$

where $x_z$ is the z-scored features of $x$ and $L$ is an $N \times K$ linear transform – as in *in* . With a purely linear code and stimuli that are defined by two features $K = 2$ that each take on two values $n = 2$, there are two stimulus pairs that give rise to exactly the same average response. Those pairs are

$$X = \{x_{11}, x_{22}\}$$
$$S = \{x_{12}, x_{21}\}$$

which is to say,

$$\bar{r}_L(X) \quad = \bar{r}_L(S)$$

Due to this property, even a decoder that is optimal for our current setting (the maximum likelihood decoder), will not be able to discriminate between these two options at an error rate different from chance. If $X$ is presented, then we refer to trials in which $S$ is decoded from the activity by a maximum likelihood decoder as a misbinding error. How can we modify the code to make fewer misbinding errors? Here, we show that reintroducing the nonlinear part of the code can effectively drive down the probability of binding errors.

Now, we return to the full code, given above. We want to show how increasing the nonlinear distance $d_N$ decreases the probability of misbinding errors. First, we derive the distance between the average representation of $X$ and $S$ in the full code (that is, with non-zero nonlinear distance), where,

$$r(X) = \sum_{x \in X} L\, x_z + M f_N(x) + \epsilon$$

So, for $X$ and $S$ defined above,

$$
\begin{aligned}
d_S \quad &= |\bar{r}(X) - \bar{r}(S)|_2 \\
&= \left| \sum_{x \in X} L\, x_z + M f_N(x) - \sum_{s \in S} L\, s_z - M f_N(s) \right|_2 \\
&= \left| \sum_{x \in X} M\, f_N(x) - \sum_{s \in S} M\, f_N(s) \right|_2 \\
&= |M_1 + M_2 - M_3 - M_4|_2 \\
&= \sqrt{2}\, d_N
\end{aligned}
$$

We notice that this does not depend on the particular stimulus sets $X$ and $S$ anymore. Indeed, due to our assumption that the nonlinear distances are constant across stimuli, this is the difference between any two sets of stimuli that are linearly confusible.

Now, knowing the distance between the correct stimulus set $X$ and a misbound stimulus set $S$, $d_S$, we can use the following expression for the rate of binding errors via a union bound approximation where $S$ remains a particular stimulus set that is linearly confusable with $X$, $C_X$ is the set of all stimulus sets that are linearly confusable with $X$, and $\hat{X}$ is the stimulus set inferred by a maximum likelihood decoder:

$$P\left(\hat{X} \in C_X | r(X)\right) = P\left(\underset{S \in C_X}{\cup} \hat{X} = S | r(X)\right)$$

$$\leq \sum_{S \in C_X} P\left(\hat{X} = S | r(X)\right)$$

$$= \sum_{S \in C_X} P\left(d\big(\bar{r}(S), r(X)\big) < d\big(\bar{r}(X), r(X)\big)\right)$$

$$\approx \sum_{S \in C_X} Q\left(-\frac{d_S}{2\sigma}\right)$$

where $d(.\, , .)$ takes the Euclidean distance between its two arguments and $Q(.)$ is the standard Gaussian cdf. Then, we average over $X$,

$$P(\text{binding error}) \approx N_S \, Q\left(-\frac{d_S}{2\sigma}\right)$$

where $N_S$ is the average size of $C_X$ across all $X$. In our case, for two stimuli with two features that each take on two values, $N_S = \frac{1}{4}$.

**Derivation of the generalization error rate**

Next, we want to develop a prediction for the generalization error rate that depends on our four parameters $d_{LV}$, $d_N$, and $\sigma$ as well as on the standard error of the mean $\epsilon$. In particular, we can compare the generalization error rate predicted by our theory and developed here from the generalization error rate of decoders trained on the neural data (discussed more in *in* ). This provides a crucial validation test for our theory, and the close correspondence observed between the theory and the data indicate that our formalization captures the relevant aspects of the geometry of the neural representations.

Here, we will develop the approximation assuming that the linear and nonlinear parts of the code are orthogonal to each other. However, the results are similar if the linear and nonlinear parts are randomly chosen with respect to each other.

As before, we have four stimuli of interest $x_{ij}$ for $i, j \in \{1,2\}$. We can write the representation corresponding to each of them in terms of the linear and nonlinear code components where $M_i$ denotes the columns of $M$ and $L_i$ denotes the columns of $L$, such that,

$$r(x_{11}) = 0$$
$$r(x_{21}) = d_{LV} L_1 + d_{N\epsilon} M_{12}$$
$$r(x_{12}) = d_{LA} L_2 + d_{N\epsilon} M_{13}$$
$$r(x_{22}) = d_{LV} L_1 + d_{LA} L_2 + d_{N\epsilon} M_{14}$$

where $d_{LA}$ is the linear distance associated with the variable that is being generalized across (i.e., offer side or time), $d_{N\epsilon}$ is the combined distance from both the nonlinear part of the code and the SEM $d_{N\epsilon} = \sqrt{d_N^2 + \epsilon^2}$, and we have defined

$$M_{ij} = \frac{M_i - M_j}{\sqrt{2}}$$

as well as for convenience. The standard error distance $\epsilon$ appears here because some of the perturbations to the underlying linear structure are reliable and form the nonlinear distance, while others are unreliable and emerge due to the noisy estimation of each centroid. The generalization performance of the classifier is reduced by both – while, for instance, the traditional cross-validated performance of the classifier would be reduced only by the standard error of the mean.

Next, we consider a linear prototype decoder – that is, a decoder that returns a binary classification of a new point by decoding whether it is closer to the class centroid for the first or second class. In this case, the two classes are separated by a linear hyperplane, which is orthogonal to the vector that connects the two class centroids (reference schematic). This method of learning the separating hyperplane does not maximize the margin between the two classes, as done by support vector machines. Here, we compute the generalization error rate for a putative prototype decoder. Later, we show that this generalization performance is a good match for the generalization performance of a support vector machine decoder.

As mentioned above, the prototype decoder uses the vector connecting the two class centroids to decode new points. In our framework, we can write an expression for this vector given the two stimuli $x_{11}$ and $x_{21}$,

$$v_D = \frac{1}{c}\big(r(x_{21}) - r(x_{11})\big)$$
$$= \frac{1}{c}(d_L L_1 + d_{N\epsilon} M_{12})$$

where $c$ normalizes $v_D$ to be a unit vector and has the form

$$c = \sqrt{d_L^2 + d_{N\epsilon}^2}$$

Because we assume that the noise magnitude is the same for each stimulus, the decoding boundary is $\frac{c}{2}$. In particular, to decode an unseen data point $y$, we would evaluate,

$$o = \text{sgn}\left[v_D \cdot y - \frac{c}{2}\right]$$

where sgn takes the sign of its argument. If $o = -1$, then we would classify $y$ as, for instance, low value; if $o = 1$, then we would classify $y$ as high value.

So, to evaluate the generalization performance of this decoder on held out stimulus conditions, we can apply this same logic to the left out stimuli $x_{12}$ and $x_{22}$. In particular, we want to project the representations of these held out stimuli onto the decoder vector $v_D$ derived above and then compare the position of the representations along that vector to the decoding threshold $\frac{c}{2}$. So, for $x_{12}$,

$$d_{12} = v_D \cdot r(x_{12}) - \frac{c}{2}$$
$$= \frac{1}{c}(d_{LV} L_1 + d_{N\epsilon} M_{12}) \cdot (d_{LA} L_2 + d_{N\epsilon} M_{13}) - \frac{c}{2}$$
$$= \frac{1}{c2} d_{N\epsilon}^2 - \frac{c}{2}$$
$$= \frac{1}{c2} d_{N\epsilon}^2 - \frac{c^2}{c2}$$
$$= \frac{d_{N\epsilon}^2}{c2} - \frac{d_{LV}^2 + d_{N\epsilon}^2}{c2}$$
$$= -\frac{\frac{1}{2} d_{LV}^2}{\sqrt{d_{LV}^2 + d_{N\epsilon}^2}}$$

and, for $x_{22}$,

$$d_{22} = v_D \cdot r(x_{22}) - \frac{c}{2}$$

$$= \frac{1}{c}(d_{LV}L_1 + d_{N\epsilon}M_{12}) \cdot (d_{LA}L_2 + d_{LV}L_1 + d_{N\epsilon}M_{14}) - \frac{c}{2}$$

$$= \frac{1}{c}\left(d_{LV}^2 + \frac{1}{2}d_{N\epsilon}^2\right) - \frac{c}{2}$$

$$= \frac{\frac{1}{2}d_{LV}^2}{\sqrt{d_{LV}^2 + d_{N\epsilon}^2}}$$

Now, using these two distances and the noise magnitude $\sigma$, we can predict how well a decoder trained to discriminate $x_{11}$ from $x_{21}$ will generalize to discriminate $x_{12}$ from $x_{22}$. This is the cross-category generalization performance from (Bernardi et al., 2020) and that is discussed in the main text. In particular, the

$$P(\text{CCGP error}) \approx \frac{1}{2}Q\left(\frac{d_{12}}{\sigma}\right) + \frac{1}{2}Q\left(-\frac{d_{22}}{\sigma}\right)$$

$$= Q\left(-\frac{1}{\sigma}\frac{\frac{1}{2}d_{LV}^2}{\sqrt{d_{LV}^2 + d_{N\epsilon}^2}}\right)$$

where $Q$ is the cumulative distribution function of a standard normal distribution, as before.

**Decoding analyses**

For decoding, we consider value presented in two distinct conditions. We begin by constructing pseudopopulations for high- and low-value in each condition (e.g., high- and low-value on the left and high- and low-value on the right). The pseudopopulation consists of all neurons from a particular brain region with at least 160 trials for each of the four conditions for the broad data splits (e.g., splitting into left and right presentations with high or low value, combining across offer 1 and offer 2) and at least 80 trials for the narrower data splits (e.g., splitting into high and low value for left offer 1 and right offer 2).

We discretize value into high and low by splitting it according to the subjective value transformation computed for each session and leaving out the middle 30 percentile. Before decoding, we preprocess the data by z-scoring and then applying a PCA that retains enough dimensions to capture 99 % of the variance. Both the z-score and PCA transforms are learned on the training set only. All decoding analyses are done on data from three non-overlapping 300 ms bins that begin 100 ms after offer onset. The activity from each neuron in each time bin are treated as separate features. All of the bins from a single trial are in either the training or testing set, they are not split across both.

Then, we train a support vector machine decoder with a linear kernel to discriminate high- from low-value in one condition (e.g., decoding value from only presentation on the left) and test that decoder both on held-out trials from that condition (10 % of trials are held out) and all trials from the other condition (e.g., high- and low-value on the right). The performance of the decoder on the held out trials from the training condition is the standard decoding performance and the performance of the decoder on the trials from the second condition is the cross-condition generalization performance.

All decoding analyses are implemented in scikit-learn (Pedregosa et al., 2011). In the main text, we compare the generalization performance of these SVM decoders to the generalization performance that is predicted by our analysis of the linear-nonlinear code model, given above. We believe that the close correspondence observed between the empirical and predicted generalization performance indicates that our linear-nonlinear code formalization captures relevant aspects of the neural representation geometry.